\documentclass{aa}
\usepackage{graphics}

\begin{document}

\thesaurus{06(08.19.4; 08.19.5 SN 1054; 09.19.2)}     

\def\EE#1{\times 10^{#1}}
\def\Msun{{~\rm M}_\odot}
\def\kms{\rm ~km~s^{-1}}
\def\lsim{\!\!\!\phantom{\le}\smash{\buildrel{}\over
  {\lower2.5dd\hbox{$\buildrel{\lower2dd\hbox{$\displaystyle<$}}\over
                               \sim$}}}\,\,}
\def\gsim{\!\!\!\phantom{\ge}\smash{\buildrel{}\over
  {\lower2.5dd\hbox{$\buildrel{\lower2dd\hbox{$\displaystyle>$}}\over
                               \sim$}}}\,\,}

\def\EE#1{\times 10^{#1}}
\def\kms{\rm ~km~s^{-1}}
\def\msunyr{~M_\odot~{\rm yr}^{-1}}
\def\Mdot{\dot M}

\title{Why did Supernova 1054 shine at late times?}

\author{J.~Sollerman,\inst{1,2}
C.~Kozma,\inst{1}
P. Lundqvist\inst{1}
}

\institute{Stockholm Observatory, SE-133 36 Saltsj\"obaden, Sweden
\and 
European Southern Observatory, Karl-Schwarzschild-Strasse 2,
D-85748 Garching bei M\"unchen, Germany}

\date{Received:  Accepted: }
\mail{jsollerm@eso.org}
\titlerunning{SN 1054 at late times}
\maketitle

\begin{abstract}
The Crab nebula is the remnant of supernova 1054 (SN 1054). 
The progenitor of this supernova has, based on
nucleosynthesis arguments, been modeled as an $8-10\Msun$ star. Here
we point out that the observations of the late light curve of SN 1054, from
the historical records, are not compatible with the standard scenario,
in  which the late time emission is powered by the radioactive decay of
small amounts of $^{56}$Ni. 
Based on model calculations we quantify this discrepancy.
The rather large mass of $^{56}$Ni needed to power the late 
time emission, $0.06^{+0.02}_{-0.03}\Msun$, 
seems inconsistent with abundances in the Crab nebula.
The late light curve may well have been powered by
the pulsar, which would
make SN 1054 unique in this respect. 
Alternatively, the late light
curve could have been powered by circumstellar interaction,
in accordance with scenarios in which $8-10\Msun$ stars are
progenitors to `dense wind' supernovae.

\keywords{supernovae: general -- supernovae: SN 1054}

\end{abstract}

\section{Introduction}

The Crab nebula is among the best studied objects in the sky. 
Still, the nature of the progenitor star and many aspects of the 
explosion remain unclear.
The ancient observations of SN 1054, conducted by 
astronomers in China and Japan, have been analyzed by Clark \&
Stephenson (1977). They conclude
that SN 1054 was observable in daytime
for 23 days and during night for some 650 days past explosion. 
Great attention has been directed to understand this light curve in terms of 
supernova theory (Minkowski 1971;  Clark \&
Stephenson 1977; Chevalier 1977; Pskovskii 1977; 
Wheeler 1978; Collins, Clapsy, \& Martin 1999, and references therein).

It is now widely accepted that SN 1054 was a core-collapse supernova, 
primarily due to the presence of the pulsar. That SN 1054 would have been a 
Type Ia supernova (SN Ia), 
an old suggestion that was recently aired by Collins et al. (1999), is
clearly in conflict also with the large amount of mass in the filaments, 
and in particular with its hydrogen-rich composition.
However, the idea that SN 1054 was a normal 
Type II supernova (SN II),
as was first suggested by Chevalier (1977), is not unproblematic.
The Crab filaments contain only $4.6\pm1.8\Msun$
of material (Fesen, Shull, \& Hurford 1997)
and cruise at merely $\sim1400$~km~s$^{-1}$ (Woltjer 1972), which give a
kinetic energy an order of magnitude less than the
canonical value for core-collapse supernovae, 10$^{51}$~ergs.
Some of this energy may also originate from the pulsar (Chevalier 1977). 
As the early observations of SN 1054 indicate a rather luminous explosion, 
Chevalier (1977) 
suggested that the missing mass and energy of the Crab resides in 
a hitherto undetected outer shell.
That some material indeed exists outside 
the visible filaments was recently shown using $\it HST$ observations 
(Sollerman et al. 2000).

Nomoto et al. (1982) constructed a detailed model for the progenitor
of SN 1054. They argued that the progenitor must have been more
massive than $8\Msun$ in order to leave a neutron star and less
massive than $\sim10\Msun$ to be consistent with the observed
metal abundances. An $8-10\Msun$ star would eject very little
heavy elements (Nomoto et al. 1982). In particular, it would eject very small 
amounts of radioactive $^{56}$Ni, responsible for the late time emission of 
the supernova (\cite{MW88}).
In this report we want to draw the attention to the long duration of the 
light curve of SN 1054, and point out that 
this cannot be explained within the standard supernova scenario, 
in which the powering of the emission at these late phases is due to 
radioactive decay of a very low mass of $^{56}$Ni.

\section{Discussion}

\subsection{Supernova Light Curves}

The early light curves of SNe II are powered by the explosion energy slowly
diffusing out of the ejecta. The large diversity of light curve shapes in
this phase (Patat et al. 1994) is largely due to variations in progenitor 
radius, envelope mass and composition, as well as 
in the explosion energy itself.
At later phases ($\gsim150$ days), the light curves of SNe II 
are often powered by 
the radioactive decay of $^{56}$Co$\rightarrow$$^{56}$Fe, 
and then the late evolution is quite uniform (Patat et al. 1994).
The $^{56}$Co is itself the
decay product of $^{56}$Ni, synthesized in the supernova explosion.
The late light curve of the well studied SN 1987A was reproduced
by models with $0.07\Msun$ of $^{56}$Ni (Kozma \& Fransson 1998a,b), 
and most SNe II light curves do follow the decay rate of $^{56}$Co
(Barbon et al. 1984; \cite{P94}). 
In fact, the luminosity on the light curve tail can be used to determine 
the mass of ejected $^{56}$Ni, as has been done for several
supernovae (see Sollerman, Cumming, \& Lundqvist 1998, and references 
therein).

The radioactive energy from the decay of $^{56}$Co
is deposited in the supernova ejecta as 
$\gamma$-rays and positrons. The luminosity on the light curve tail from the
$\gamma$-rays is given by 
$L_\gamma=
1.26\times10^{43}F_\gamma~M_{\rm Ni}$~e$^{-t/111.3}$~ergs~s$^{-1}$,
where $F_\gamma$ is the fraction of the $\gamma$-rays trapped in the ejecta, 
$M_{\rm Ni}$ is the amount of $^{56}$Ni in solar masses, and 111.3 days
is the e-folding time for the decay of $^{56}$Co.
The positrons contribute 
$L_{e^{+}}=4.45\times10^{41}~M_{\rm Ni}$~e$^{-t/111.3}$~ergs~s$^{-1}$, 
assuming they are all deposited locally.

To determine the nickel mass from the observed filter light curves, 
the $\gamma$-ray trapping and the bolometric evolution have to be taken into 
account. In Figure 1 we use SN 1987A to illustrate these concepts. 
There we plot the bolometric light curve of SN 1987A (triangles) from 
Bouchet \& Danziger (1993).
The upper dashed line is given by the expressions above
for a nickel mass of $0.07\Msun$ (\cite{SB90}) assuming full trapping
of the $\gamma$-rays ($F_\gamma$=1). This line fits the
light curve tail at epochs up to $\sim300$ days, but later the
observed luminosity falls below the luminosity expected for
full trapping, due to an increasing leakage of $\gamma$-rays.

   \begin{figure}[!ht]
\begin{center}
\resizebox{\hsize}{!}{\includegraphics{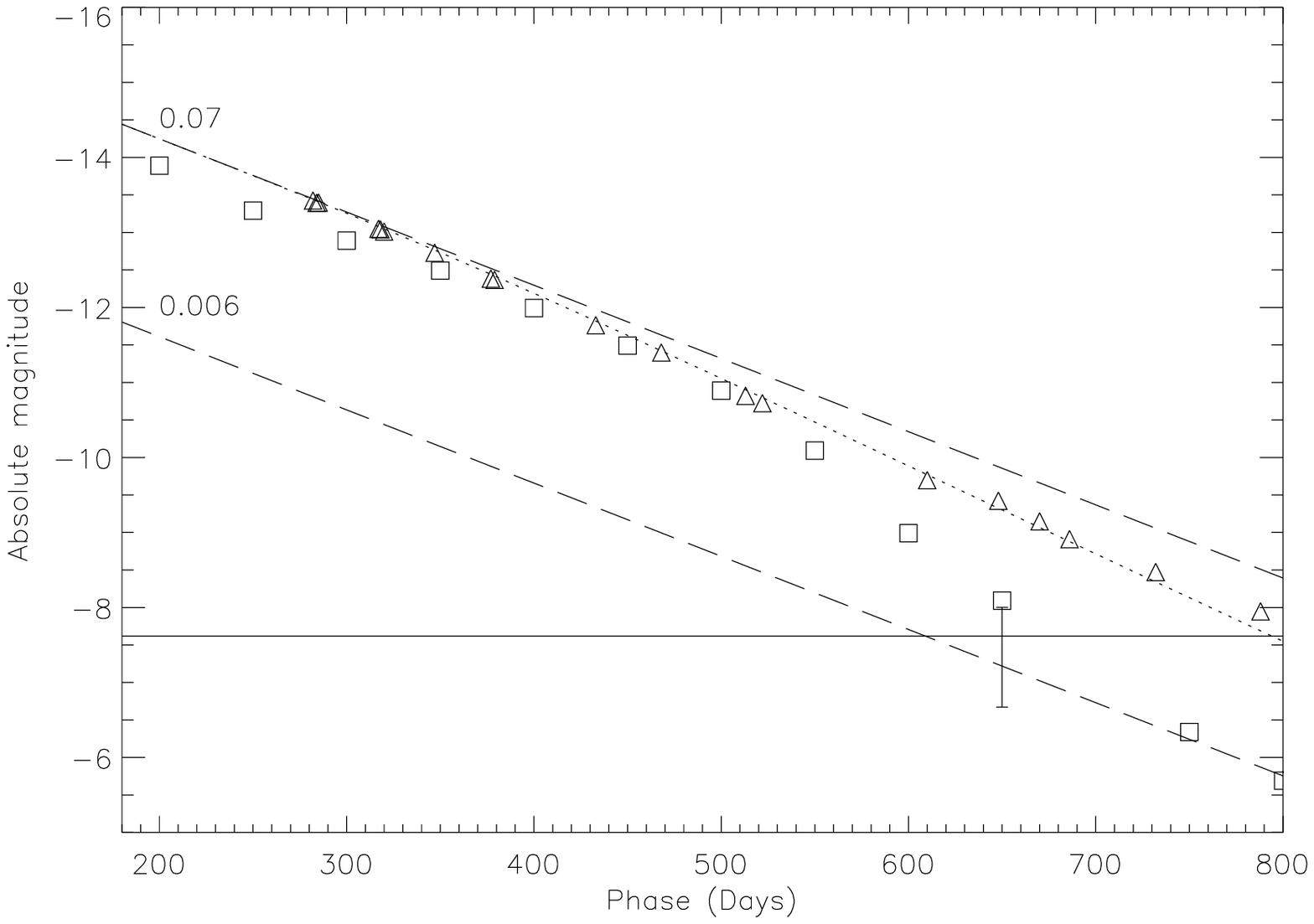}}
      \caption[]{

The upper long dashed line labeled `0.07' gives the absolute 
bolometric magnitude expected from the radioactive decay of $0.07\Msun$ of
$^{56}$Ni, assuming full trapping. The dotted line allows for
$\gamma$-ray leakage with $t_{1}=610$, as explained in the text. 
This gives a decent fit to the bolometric magnitudes 
of SN 1987A (50 kpc, $E(B-V)=0.16$) 
shown by the triangles (Bouchet \& Danziger 1993).
The square symbols show the absolute $V$-band magnitudes for SN 1987A 
(Suntzeff \& Bouchet 1990), which deviate substantially from the bolometric 
curve at late phases.
The lower dashed line is the absolute bolometric magnitude
expected from the radioactive decay of $0.006\Msun$
$^{56}$Ni, again assuming full trapping.
The horizontal line represents the naked eye detection limit for a
source at a distance of 2.0 kpc and with an extinction of $E(B-V)=0.52$, the
parameters for SN 1054. The error bar shown at day 650 for this
detection limit encapsulates the errors described in the text.
}
         \label{Fig1}
\end{center}
   \end{figure}

The $\gamma$-leakage can be illustrated using a simple model with 
a central radioactive source, where 
the deposition $F_\gamma$=(1~$-$~e$^{-\tau}$), and
the $\gamma$-ray optical depth evolves as
$\tau=(t/t_{1})^{-2}$ due to the homologous expansion. 
Here $t_{1}$ is the time when $\tau$=1.
For SN 1987A, $t_{1}\sim610$ days gives a reasonable agreement with the
bolometric magnitudes 
as shown by the dotted curve in Figure 1.
This means that $\sim60\%$ of the
$\gamma$-rays were trapped at 650 days for SN 1987A.

Also plotted in Figure 1 is the $V$-band light curve of SN 1987A
(square symbols, \cite{SB90}). It can be seen that this filter 
light curve deviates
substantially from the bolometric light curve after about 500
days. At these late phases, the ejecta temperature is low and most of
the light is instead emitted in the infrared (Kozma \& Fransson 1998a,b).

\subsection{The late time luminosity of SN 1054}

The Crab is located at a distance of about 2.0 kpc 
(\cite{T73}) with an extinction of $E(B-V)=0.52$ (Sollerman et al. 2000). 
From the historical records analyzed by Clark \& Stephenson (1977) we
know that SN 1054 faded from visibility some 650 days after discovery. 
The detection limit for night time observations was estimated to be 
5.5 mag by these authors (Clark \& Stephenson 1977).
SN 1054 must thus have had an absolute $V$ (actually visual) 
magnitude of M$_V=-7.6$ at this epoch. 
In Figure 1 we have indicated this limit together with an 
error bar that encapsulates distances in the range 
$1500-2200$ pc (\cite{DF85}), and
an error of $\pm0.04$ in $E(B-V)$. As a detection limit of 5.5
magnitudes may be regarded too high, we encapsulate it with 
conservative limits of $5.5^{+0.7}_{-0.3}$. This is also included in the
error bar in Figure 1.

The Crab progenitor has been modeled as an $8-10\Msun$ star (\cite{N82}).
The amount of $^{56}$Ni ejected in such an explosion 
is supposed to be very small. Detailed
calculations performed by Mayle \& Wilson (1988)
indicate that no more than $0.002\Msun$ of $^{56}$Ni
should be ejected from supernovae with progenitors in this mass range. 

The $^{56}$Ni eventually decays to $^{56}$Fe 
and thus the current amount of iron probes the mass of ejected
nickel. Abundance analyses in the Crab 
are nontrivial, but indicate a subsolar iron abundance 
(Davidson 1978,1979; \cite{H84}; \cite{BPP96}).
For an ejecta mass of $4.6\Msun$, solar abundance corresponds to 
$0.006\Msun$ of iron. This is consistent with a low mass of ejected
nickel, as suggested by the explosion models of Mayle \& Wilson (1988).

Here we simply want to point out that the naked eye observations of SN 1054
at 650 days seem to be inconsistent with
the standard scenario where supernovae
from $8-10\Msun$ stars are powered at late times by the
radioactive decay of very small amounts of $^{56}$Ni.

Even assuming full trapping
and that all of the emitted flux emerges in the visual band,
(F$_\gamma$=1, M$_{V}$=M$_{bol}$), 
just barely keeps the supernova shining at 650 days past explosion. 
The lower dashed line labeled `0.006' in Figure 1 shows a 
full trapping, bolometric case for this amount of ejected $^{56}$Ni,
and is fairly close to the lower limit of 
the luminosity needed for naked eye visibility of 
the supernova at 650 days. 

This discrepancy is of course much 
more pronounced in the more realistic case when
the conservative assumptions above are relaxed.
For example, assuming that SN 1054 had the same $\gamma$-ray leakage 
and bolometric correction as SN 1987A, $\sim0.05\Msun$ 
of $^{56}$Ni would be required to reach M$_V=-7.6$. 

\subsection{Modeling}

Clearly, it is too simplistic to directly compare SN 1054 with SN 1987A. 
To quantify the discrepancy mentioned above, we have therefore 
modeled SN 1054 in some detail.
Lacking an explosion model, we 
base our input model on the available observations. 
We have adopted an ejecta mass of $4.6\Msun$ (\cite{F97}) 
and a maximum velocity of 2300~km~s$^{-1}$ (Clark et al. 1983; \cite{DF85}).
We assume that the density and composition are constant throughout the ejecta.
We use solar abundances from Cameron (1982)
except for helium, for which we use 
a ratio of the mass fractions of helium and 
hydrogen, X(He)/X(H)=2 (e.g., Henry 1986). 

A central source of radioactive $^{56}$Ni is assumed. 
Calculations 
have been done for $0.006\Msun$, $0.04\Msun$  and $0.07\Msun$ of $^{56}$Ni.
The emission from SN 1054 is modeled in detail 
from 200 to 800 days. The code is fully
described in Kozma \& Fransson (1998a,b).
In the decay of 
$^{56}$Co, $\gamma$-rays and positrons are emitted, and the 
thermalization of these is 
calculated using the Spencer-Fano formalism (Kozma \& Fransson 1992).
This time-dependent code successfully reproduces the detailed late time 
observations of SN 1987A (Kozma 2000).

   \begin{figure}[!ht]
\begin{center}
\resizebox{\hsize}{!}{\includegraphics{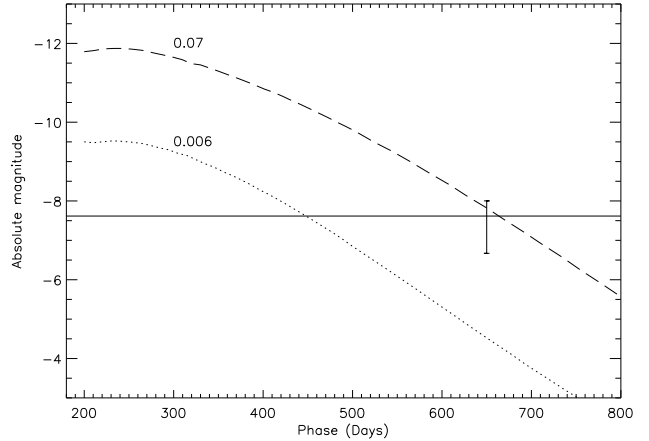}}
      \caption[]{
Absolute 
$V$-band light curves for the Crab model. The upper curve is for $0.07\Msun$
of $^{56}$Ni, while the lower curve is for $0.006\Msun$ of $^{56}$Ni.
The detection limit is the same as in Fig. 1.
}
         \label{Fig2}
\end{center}
   \end{figure}

With this model, we estimate that
only $32\%$ of the $\gamma$-rays were trapped at 650 days for SN 1054. 
The low trapping 
is of course due to the low mass of the ejecta, as compared to SN 1987A.
In Figure 2 we plot the light curves for the SN 1054 model 
for two different amounts of ejected $^{56}$Ni. 
At 650 days, a smaller amount of the luminosity comes out in the $V$-band for 
the low nickel-mass model.
This is because low amounts of nickel means less heating, 
and as the temperature of the ejecta decreases the emission is pushed further 
into the infrared.

From Figure 2 we directly see that $0.006\Msun$ of $^{56}$Ni
could not have provided the observed
luminosity at 650 days. Instead, the model with $0.07\Msun$ of 
$^{56}$Ni (similar to the mass ejected in SN 1987A), comes 
close to a $V$-magnitude of 5.5 at day 650.
Within the framework of this model, a mass of $^{56}$Ni
of $0.06^{+0.02}_{-0.03}\Msun$ was required to power the 
supernova light curve when it faded from night time visibility.
This includes the errors estimated above for distance, reddening and 
naked eye-sensitivity.
We have done the same calculations also for a filter similar to the eye 
sensitivity (Rhodopsin absorption curve, Kitchin 1991), but the deviation 
from the $V$-filter is small, 0.07 mag for the model with 
$0.07\Msun$ of $^{56}$Ni at 650 days.
Finally, although according to Clark \& Stephenson (1977) there is no
reason to question the dates of the Chinese sightings, 
an error of $\pm20$ days in duration would introduce a $30\%$ error in flux.
We note that suggestions for alternative durations, although
often based on less direct sources (e.g., Pskovskii 1977; 
Collins et al. 1999), usually argue for longer visibility of SN 1054,
thus enhancing the problem of keeping it shining at late times.

\section{Powering scenarios}

\subsection{Radioactive decay}

If radioactivity alone powered the late time emission, 
the required mass of $^{56}$Ni
estimated above, $0.06^{+0.02}_{-0.03}\Msun$, is  
significantly higher than
that obtained in the explosion models by Mayle \& Wilson (1988), 
$0.002\Msun$.
Although the unknown mechanism of core-collapse explosions makes
the existing explosion models rather uncertain,
including the exact amount of nickel mass, 
there are other reasons to believe that low mass SNe eject very small 
amounts of $^{56}$Ni. 
Galactic chemical evolution models imply that the amount of iron produced by
SNe II is already quite high, and Samland (1998) 
even suggested that the lower mass limit of
stars producing iron should be 
$11-12\Msun$, instead of the
conventional $10\Msun$. 

Nevertheless, the above arguments cannot conclusively argue that SN 1054
did not eject the $0.06\Msun$ of $^{56}$Ni required to maintain the late 
time luminosity. We therefore turn the attention to the measured abundances 
in the Crab nebula. 

Actually, only helium is enhanced in this supernova 
remnant, while the abundances of C, N, O and Fe seem to be (sub)solar 
(\cite{DF85}). In \S 2.2 we saw that solar abundance of iron for 
$4.6\Msun$ ejecta corresponds to $0.006\Msun$. 
The amount of $^{56}$Ni required to power the late time light curve,
$0.06^{+0.02}_{-0.03}\Msun$, would thus correspond to a  
present iron abundance 
which is $9^{+4}_{-5}$ times higher than solar abundance. 

If radioactivity powered the supernova, the
iron must be locked up in dust.
Dust is clearly present in the nebula, and it is nontrivial to determine the 
dust mass (Sankrit et al. 1998), although 
estimates seem to indicate that the dust mass is low
(\cite{DF85}).
Note that this scenario would also require large amounts of other metals 
to be locked up in dust. 
As the iron-group elements are produced in the very center of the 
exploding star, we do not expect to find a large amount of iron, 
while the CNO-abundance has not been enhanced.
For example, SN 1987A, which ejected $0.07\Msun$ of $^{56}$Ni,
also ejected almost 2 solar masses of oxygen. 
Even the $11\Msun$ model of Woosley \& Weaver (1995), which
ejects $0.07\Msun$ of $^{56}$Ni, and only $0.136\Msun$ of oxygen,
provides an oxygen abundance higher than solar.
We believe that the scenario in which SN 1054 was powered by 
$0.06\Msun$ of $^{56}$Ni, but where all the nucleosynthesized 
metals are presently locked up in dust, is not very convincing. We therefore 
briefly turn our attention to alternative scenarios.

\subsection{The Pulsar}

If radioactivity did not power SN 1054 at late times, 
the obvious candidate is the pulsar that
powers the nebula today. Already 
Chevalier (1977) suggested that the pulsar could
contribute to the late time supernova light curve. 
In fact, unless the bulk of the iron is locked up in dust, 
and circumstellar interaction was unimportant (see \S 3.3),
the pulsar would have to make up for all
of the late time luminosity, as $0.006\Msun$ of $^{56}$Ni
could not have kept the SN visible for more than $\sim500$ days (Fig. 2).

There are many ways in which a pulsar can contribute to the optical
luminosity of a supernova. Accretion onto the neutron star could be
either spherical or in a disk, depending on the angular momentum of the
infalling matter. A steady spherical accretion is supposed to reach a
maximum Eddington luminosity of 
$L_{\rm Edd}=3.5\times10^{38}$
ergs~s$^{-1}$, and 
Chevalier (1989) suggested that this scenario might be responsible for
the late time luminosity of SN 1054. He worried, however, about the
fact that the accretion luminosity would not be able to escape from
the vicinity of the neutron star during the first months, as the
luminosity is trapped by the inflow (see also Benetti et al. 2000).
As shown in Figure 2, this need not be a problem, as even a
low mass of ejected $^{56}$Ni is able to keep the supernova
shining for some 500 days. If the accretion is mediated via a
disk, the luminosity could perhaps be even larger than in the spherical case.

Unknown in these scenarios is the fraction of the luminosity
escaping in different bands. In fact, the full Eddington luminosity
would have to fall in the visible band to make SN 1054 observable at 650
days ($3.5\times10^{38}$~ergs~s$^{-1}$ corresponds to M$_{\rm bol}=-7.7$).
This appears to be rather unlikely. 

Pulsar nebulae in supernovae, with special attention to the
Crab, were investigated by Chevalier \& Fransson (1992).
Unfortunately, they did not address the question highlighted here, 
the luminosity of SN 1054 at 650 days. 
In their scenario, the pulsar powers a bubble that shocks and ionizes
the supernova ejecta, and their calculations
suggested that $1.5\%$ of the total pulsar luminosity could be converted
to radiation. This would be too low to account for the
luminosity of SN 1054 at day 650. However, 
if the pulsar bubble is a significant source of synchrotron
emission (Chevalier 1996), 
the efficiency might be higher. 
Another interesting possibility is that the pulsar was born with a very rapid 
spin period
(see Atoyan 1999), and thus a very high initial spin-down luminosity.

In summary, although not investigated in detail,
there are several ways in which a pulsar could 
contribute to the luminosity of supernovae. 
This has so far never been unambiguously observed. 
If the pulsar was indeed responsible for the late light curve, 
this would make SN 1054 a unique case.

\subsection{Circumstellar interaction}

Another mechanism important for supernova light curves is interaction
with circumstellar material (CSM). 
Circumstellar interaction is responsible for the emission at late phases 
for a number of supernovae, such as SNe 1979C, 1980K, 1988Z, 1993J and 1995N.
Such interaction can maintain a SN luminosity of several$\times 10^{38}$
ergs~s$^{-1}$ for many years (e.g., Chugai, Danziger, \& Della Valle 1995).
In this respect, 
we note that the Crab could fit into the scenario of $8-10\Msun$ stars 
being progenitors to the so called `dense wind' supernovae 
(\cite{C97}), where the supernova ejecta interact with a 
dense superwind from the progenitor star. 

The dense wind must have extended out to $\gsim 6\EE{16}$ cm
for a maximum ejecta velocity of $v_{\rm ej} \sim10^4 \kms$.
This means that the wind 
started $\gsim 2 (v_{\rm ej}/v_{\rm w})$ years before the SN
breakout. Here $v_{\rm w}$ is the wind velocity.
Comparing with the line fluxes in 
Chevalier \& Fransson (1994), we find that the late
light curve can be explained if $\Mdot / v_{\rm w}$ 
was typically in excess of $10^{-5} \msunyr / 10 \kms$.  
Here $\Mdot$ is the mass loss
rate during the superwind. The swept up wind should 
in this case now coast freely outside the present nebula.
However, no sign of ejecta, or swept up wind, moving at velocities of
order $10^4 \kms$ has been identified (e.g., Fesen et al. 1997).

Chugai \& Utrobin (1999) suggested that SN 1054 was 
similar to the low-energy ($\sim 4\EE{50}$ erg) event 
SN 1997D (Turatto et al. 1998; Benetti et al. 2000),
and that no fast ejecta exist. In this case, circumstellar powering
of the late light curve could be more difficult as the luminosity of
the circumstellar shock scales as $\propto v_{\rm ej}^3$. 
Furthermore, in a low-energy explosion, the peak luminosity is more 
likely to be lower, and Chugai \& 
Utrobin (1999) therefore suggested that circumstellar 
interaction was important for SN 1054 also at the early phase. A low-energy 
explosion model with a circumstellar shell was explored already 
by Falk \& Arnett (1977), and they found a high efficiency in 
the conversion of shock energy to visual light during the peak,
in accordance with the suggestion of Chugai \& Utrobin (1999).    
Although there is thus no direct support for circumstellar interaction 
as the cause for the late emission in the Crab, it is still premature
to rule out this possibility.

\section{Summary}
SN 1054, the creation of the Crab nebula and the Crab pulsar, is the
typical example of an $8-10\Msun$ supernova.
Such supernovae are expected to eject only
minute amounts of $^{56}$Ni, and here we demonstrate
that SN 1054 could not have been powered by such a small mass
of nickel at late times. The required amount of
$^{56}$Ni, $\sim0.06\Msun$, is much larger than suggested by
abundance analyses of the Crab, and we therefore discuss
alternative solutions.
The pulsar may have powered the supernova, which would make SN 1054
unique in this respect. Alternatively, the progenitor of SN 1054 could
have had a dense wind and the supernova could then have been 
powered by circum-stellar interaction.

\begin{acknowledgements}
We thank the referee for careful reading.
\end{acknowledgements}


\end{document}